\documentclass[conference]{IEEEtran}
\IEEEoverridecommandlockouts
\usepackage{cite}
\usepackage{amsmath,amssymb,amsfonts}
\usepackage{algorithmic}
\usepackage{graphicx}
\usepackage{textcomp}
\usepackage{xcolor}
\usepackage{url}
\usepackage{subcaption}
\usepackage{makecell}

\def\BibTeX{{\rm B\kern-.05em{\sc i\kern-.025em b}\kern-.08em
    T\kern-.1667em\lower.7ex\hbox{E}\kern-.125emX}}

\begin{document}

\title{QTRL: Toward Practical Quantum Reinforcement Learning via Quantum-Train}

\author{
\IEEEauthorblockN{
    Chen-Yu Liu \IEEEauthorrefmark{1}\IEEEauthorrefmark{2}\IEEEauthorrefmark{6}, 
    Chu-Hsuan Abraham Lin\IEEEauthorrefmark{3}\IEEEauthorrefmark{7}, \\Chao-Han Huck Yang
    \IEEEauthorrefmark{9}\IEEEauthorrefmark{11},
    Kuan-Cheng Chen\IEEEauthorrefmark{4}\IEEEauthorrefmark{5}\IEEEauthorrefmark{8},
    Min-Hsiu Hsieh
    \IEEEauthorrefmark{1}
    \IEEEauthorrefmark{10}
}
\IEEEauthorblockA{\IEEEauthorrefmark{1} Hon Hai (Foxconn) Research Institute, Taipei, Taiwan}
\IEEEauthorblockA{\IEEEauthorrefmark{2}Graduate Institute of Applied Physics, National Taiwan University, Taipei, Taiwan}
\IEEEauthorblockA{\IEEEauthorrefmark{3}Department of Electrical and Electronic Engineering, Imperial College London, London, UK}
\IEEEauthorblockA{\IEEEauthorrefmark{4}Department of Materials, Imperial College London, London, UK}
\IEEEauthorblockA{\IEEEauthorrefmark{5}Centre for Quantum Engineering, Science and Technology (QuEST), Imperial College London, London, UK}
\IEEEauthorblockA{\IEEEauthorrefmark{9}NVIDIA, Taipei, Taiwan}
\IEEEauthorblockA{Email:
\IEEEauthorrefmark{6}
chen-yu.liu@foxconn.com, 
\IEEEauthorrefmark{7} abraham.lin23@imperial.ac.uk, \\ \IEEEauthorrefmark{8}
kuan-cheng.chen17@imperial.ac.uk,
\IEEEauthorrefmark{11}
hucky@nvidia.com,
\IEEEauthorrefmark{10}
min-hsiu.hsieh@foxconn.com
}}

\maketitle

\begin{abstract}
Quantum reinforcement learning utilizes quantum layers to process information within a machine learning model. However, both pure and hybrid quantum reinforcement learning face challenges such as data encoding and the use of quantum computers during the inference stage. We apply the Quantum-Train method to reinforcement learning tasks, called QTRL, training the classical policy network model using a quantum machine learning model with polylogarithmic parameter reduction. This QTRL approach eliminates the data encoding issues of conventional quantum machine learning and reduces the training parameters of the corresponding classical policy network. Most importantly, the training result of the QTRL is a classical model, meaning the inference stage only requires classical computer. This is extremely practical and cost-efficient for reinforcement learning tasks, where low-latency feedback from the policy model is essential.
\end{abstract}


\section{Introduction}
Quantum machine learning (QML) is rapidly emerging as a transformative field. By harnessing the unique computational capabilities of quantum mechanics, such as quantum superposition and entanglement, QML aims to revolutionize neural network (NN) training and performance \cite{qml1, qml2, qml3, qml4}. Integrating Grover’s search algorithm with QML, particularly for classification tasks, promises substantial performance enhancements. From an application perspective, QML exhibits immense potential in tackling complex datasets, with applications ranging from drug discovery and large-scale stellar classification to natural language processing, recommendation systems, and generative learning models \cite{gs1, gsml1, qmlapp1, qmlapp2, qmlapp3, qmlapp4, qmlapp5, qnlp1, qrs1, qrs2, qgan1, qgan2, qgan3, qgan4, qgan5}.

Similar to machine learning (ML) and QML, quantum reinforcement learning (QRL) \cite{rl1, qrl1, qrl2} integrates quantum layers and has the potential to solve complex decision-making problems more effectively than classical approaches. However, QRL faces significant challenges, particularly in the areas of data encoding and practical implementation. Encoding classical data into quantum states efficiently is a non-trivial task, often requiring sophisticated techniques that can introduce overhead and complexity \cite{dataru1, qtl1, qrac1}. Furthermore, the practical deployment of QRL systems is hindered by the need for quantum hardware during both training and inference stages, which can be costly and less accessible than classical approaches. Addressing these challenges is crucial for realizing the full potential of QRL and making it a viable option for a wide range of applications.


The Quantum-Train (QT) approach \cite{qt1, qt2}, on the other hand, uses a quantum neural network (QNN) combined with a mapping model to generate the parameters of a classical machine learning model. By leveraging the fact that an $n$-qubit quantum state has a Hilbert space of size $2^n$, we can generate up to $2^n$ distinct measurement probabilities corresponding to the measured basis. With a polynomial number of layers with respect to the number of qubits, it is possible to control $2^n$ parameters using only $O(\text{poly}(n))$ rotational angles in the QNN, as we will describe in more detail in a later section. Consequently, the training result of QT using the quantum approach is a classical neural network model. When applying this method to the policy model of a policy gradient reinforcement learning task, this Quantum-Train reinforcement learning (QTRL) framework not only eliminates the data encoding issue of the QNN but also makes the trained result independent of the usage of quantum hardware. 

\section{Quantum-Train as Policy Network Parameter Generator}
We will start by introducing the policy gradient method, the reinforcement learning approach used in examining the QTRL framework. This will be followed by a description of the QT approach and an explanation of how we combine QT and RL.

\subsection{Policy Gradient Method}

The policy gradient method is a well-established approach in RL that focuses on training agents to make decisions through interaction with an environment \cite{rl1, rl2}. An agent aims to perform actions that maximize cumulative rewards over time. This process is typically modeled as a Markov Decision Process (MDP), which comprises a state space $\mathbb{S}$, an action space $\mathbb{A}$, a reward function, and transition probabilities. The primary objective is to learn an optimal policy $\pi^*(a|s)$, which prescribes the best action $a \in \mathbb{A}$ to take in each state $s \in \mathbb{S}$, thereby maximizing the expected discounted return:

\begin{equation}
\pi^* = \arg\max_{\pi} \mathbb{E}_{\pi} \left[ \sum_{t=0}^{T} \gamma^t R_t \right],
\end{equation}

where $R_t$ is the reward obtained at time step $t$, and $\gamma \in [0, 1]$ is a discount factor that determines the relative importance of immediate versus future rewards.

To adjust the policy, $\pi(a|s; \theta)$ is typically parameterized by a neural network model with parameters $\theta$, where an input $s$ produces an output $a$. The goal is to optimize $\theta$ to maximize the expected return. The fundamental principle involves using gradient to update $\theta$ in a manner that increases the probability of actions leading to higher rewards. This optimization process is crucial for improving the agent’s decision-making capabilities. To achieve this, the policy loss function is then defined as the negative log probability of the taken actions, weighted by the normalized returns $R^{\prime}_t$:
\begin{equation}
\mathcal{L}(\theta) = - \sum_{t=0}^{T-1} \log \pi(a_t | s_t; \theta) \cdot R^{\prime}_t,
\end{equation}
where $R^{\prime}_t = \frac{R_t - \mu_R}{\sigma_R}$, $\mu_R$ is the mean of the returns and $\sigma_R$ is the standard deviation. This loss function encourages the policy to increase the probability of actions that lead to higher returns. The gradient of the loss function with respect to the policy parameters $\theta$ is used to update the policy network:
\begin{equation}
\label{eq:gradient}
\nabla_\theta \mathcal{L} = - \sum_{t=0}^{T-1} \nabla_\theta \log \pi(a_t | s_t; \theta) \cdot R^{\prime}_t.
\end{equation}
By computing this gradient, the policy parameters $\theta$ can be adjusted to improve the policy. 

\subsection{Quantum-Train}

\begin{figure*}[ht]
\centering
\includegraphics[scale=0.27]{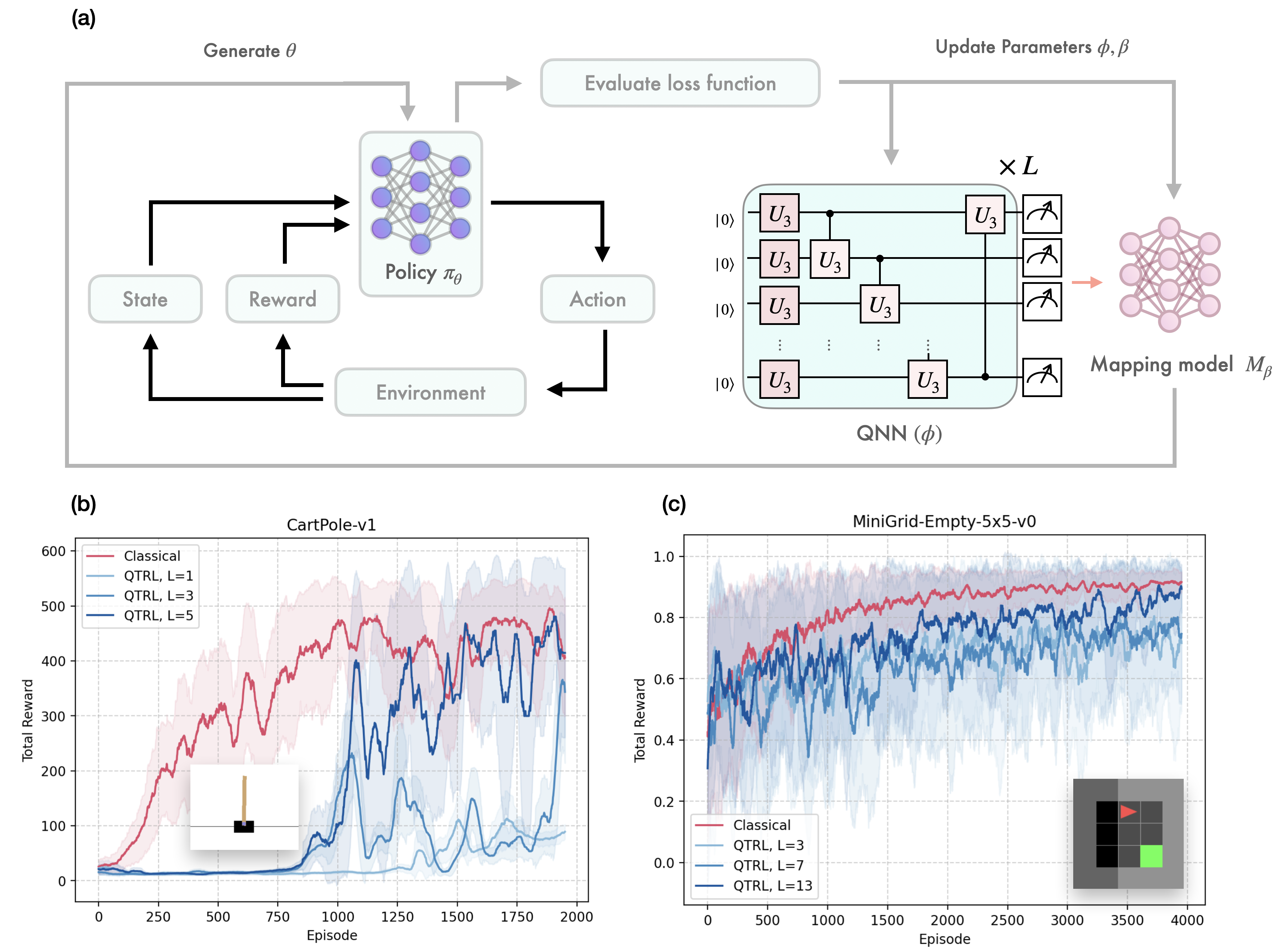}
\caption{(a) Illustration of QTRL framework. (b) Total reward as a function of episode number for the \textsf{\small CartPole-v1} environment. The classical method is shown in red, while QTRL with varying depths $(L = 1, 3, 5)$ are shown in different shades of blue. (c) Total reward as a function of episode number for the \textsf{\small MiniGrid-Empty-5x5-v0} environment. The classical method is shown in red, while QTRL with varying depths $(L = 3, 7, 13)$ are shown in different shades of blue. Insets show the visual representation of each environment.}
\label{fig:scheme}
\end{figure*}

Suppose the neural network model with parameters $\theta$ has $k$ trainable parameters: $\theta = (\theta_1, \theta_2, \ldots, \theta_k)$. In conventional ML and RL training, it is necessary to initialize $k$ parameters and tune all $k$ parameters during the training process. In contrast, the QT framework \cite{qt1, qt2} utilizes a parameterized quantum state $| \psi (\phi) \rangle$ with $n = \lceil \log_2 k \rceil$ qubits and parameters $\phi$, represented by a QNN, to generate $2^n \ge k$ distinct measurement probabilities $|\langle i | \psi (\phi) \rangle|^2$ for $i \in \{1, 2, \ldots, 2^n\}$. Using a mapping model $M_{\beta}$, which is an additional classical neural network with parameters $\beta$, the first $k$ basis measurement probabilities are mapped from values bounded between 0 and 1 to $-\infty$ and $\infty$ as follows:
\begin{equation}
\label{eq:mapping}
    M_{\beta}(i_b, |\langle i | \psi (\phi) \rangle|^2) = \theta_i, \quad i = 1, 2, \ldots, k
\end{equation}
where $i_b$ is the bit-string representation of the basis $|i\rangle$. Thus, $\theta$ is generated from the output of the QNN and the mapping model $M_\beta$. Consequently, tuning $\phi$ and $\beta$ effectively changes the value of the loss function. With the $\beta$ and $\phi$ dependency of $\theta$ from Eq.~(\ref{eq:mapping}), $\theta = M_{\beta}(\phi)$, the gradient of the loss function with respect to $\phi$ and $\beta$ can be derived as follows:
\begin{equation}
\label{eq:gradient}
\nabla_{\beta, \phi} \mathcal{L} = - \sum_{t=0}^{T-1} \nabla_{\beta, \phi} \log \pi(a_t | s_t; \beta, \phi) \cdot R^{\prime}_t.
\end{equation}
Thus, it is possible to train the policy by only tuning $\phi$ and $\beta$. As for the number of parameters in $\phi$ and $\beta$, and their relation to $k$, suppose that the mapping in Eq.~(\ref{eq:mapping}) could be constructed with polynomial depth of the QNN. Since $n = \lceil \log_2 k \rceil$, the required number of parameters for $\phi$ is $O(\text{polylog}(k))$. Given that the input to $M_\beta$ is the bit-string representation of the basis with length $n$ and a scalar, the combined vector of length $n+1$, with a polynomial depth neural network, could also require $O(\text{polylog}(k))$ parameters. Therefore, $\theta$, with the number of parameters $k$, is generated by $O(\text{polylog}(k))$ parameters. The graphical illustration of the QTRL framework is depicted in Fig.~\ref{fig:scheme}(a). 

\subsection{Natural Policy Gradient Convergence on QTRL}

\textbf{Quantum Influence on Gradient Estimation.}
The parameters $\theta$ of the policy network are generated by a QNN combined with a mapping model. These quantum-related parameters, denoted as $(\phi, \beta)$, influence the policy network parameters through the quantum state preparation and measurement processes. The gradient of the loss function, reflecting the quantum context, is given by:
\begin{equation}
\nabla_{\beta,\phi} \mathcal{L}(\pi_\theta) = \left( \frac{\partial \theta}{\partial (\beta,\phi)} \right)^T \cdot \nabla_{\theta} \mathcal{L}(\pi_\theta)
\end{equation}

where $\frac{\partial \theta}{\partial (\beta,\phi)}$ is the Jacobian matrix that represents the sensitivity of classical parameters $\theta$ to changes in the quantum parameters $(\phi, \beta)$.

\textbf{Learning Rate and Update Rule.}
The learning rate $\eta$ plays a crucial role, especially given the complex dynamics introduced by the quantum-classical interface. The update for the quantum parameters can be modeled as:
\begin{equation}
\phi_{t+1}, \beta_{t+1} = \phi_t, \beta_t + \eta \nabla_{\phi, \beta} \mathcal{L}(\pi_{\theta_t})
\end{equation}

This update rule ensures that the quantum parameters are adjusted to optimize the performance of the policy gradient.

\textbf{Regret Decomposition.}
Utilizing the performance difference lemma and the Lipschitz continuity of the value function, we can express the improvement in policy performance from one iteration to the next as:
\begin{equation}
\mathcal{L}(\pi_{t+1}) - \mathcal{L}(\pi_t) \geq \gamma \langle \nabla_{\theta} \mathcal{L}(\pi_{\theta_t}), \Delta \theta_t \rangle - \frac{L}{2} \|\Delta \theta_t\|^2
\end{equation}
Here, $\gamma$ is a factor derived from policy improvement steps, and $L$ is a Lipschitz constant for the value function, reflecting the smoothness in policy updates.

\textbf{Expected Regret Bound.}
The expected regret after $T$ steps can be bounded using the aggregation of differences over all timesteps. Considering the telescoping nature of these sums, the expected regret is estimated as:
\begin{equation}
\text{Regret}_T \leq \frac{1}{\gamma} \left( \mathcal{L}(\pi^*) - \mathcal{L}(\pi_1) \right) + \frac{L}{2} \sum_{t=1}^T \|\Delta \theta_t\|^2
\label{ref:reg}
\end{equation}
The expression in Eq.~(\ref{ref:reg}) shows the maximum expected regret in terms of the initial suboptimality and the sum of squared updates, highlighting the influence of both the learning rate and the stability of the quantum parameter generation process.

This theoretical framework provides an approach to adapting regret analysis for NPG when the policy parameters are influenced by quantum-generated parameters. Further experimental and empirical studies are necessary to refine these theoretical predictions, ensuring that quantum computational benefits translate into measurable improvements in policy performance.

In our setup, we utilize a specific type of quantum gate known as the $U_3$ gate for the QNN. This gate is crucial for our purposes because it allows us to adjust the quantum state in very precise ways, characterized by its matrix representation:
\begin{equation}
    U_3(\mu, \varphi, \lambda) = \left[ \begin{array}{cc}
    \cos(\mu/2) & -e^{i \lambda} \sin(\mu/2) \\
    e^{i \varphi} \sin(\mu/2) & e^{i(\varphi + \lambda)} \cos(\mu/2)
    \end{array} \right]
\end{equation}
In conjunction with the $U_3$ gate, the controlled-$U_3$ gate (or $CU_3$) plays a crucial role to entangle qubits:
\begin{equation}
    CU_3 = I \otimes |0\rangle \langle 0 | + U_3(\mu, \varphi, \lambda) \otimes |1\rangle \langle 1 |
\end{equation}
These parameterized gates, particularly the $CU_3$ with its circular layout, enable the specification of the number of parameters as a polynomial function of the qubit count.

\section{Result and Discussion}

To examine the applicability of the QTRL approach, we tested it in two well-known environments: \textsf{\small CartPole-v1} \cite{cartpole} and \textsf{\small MiniGrid-Empty-5x5-v0} \cite{minigrid}. The quantum circuit simulations were conducted using the TorchQuantum package \cite{torchquantum}, where the quantum states involved were simulated using state vector simulation.

\begin{table*}[!htp]
\centering
\begin{tabular}{|l|l|c|c|}
\hline
\textbf{Model} & \textbf{Topology} & \makecell{\textbf{Last 10 episode} \\ \textbf{average reward}} & \textbf{Para. size} \\
\hline
Classical Policy Model & (4-128, 128-2) & 436.3 & 898 \\
\hline
QTRL-1 & QNN: ($U_3$ block)*1 + Mapping Model: (11-10, 10-10, 10-1) & 94.6 & 361 \\
QTRL-3 & QNN: ($U_3$ block)*3 + Mapping Model: (11-10, 10-10, 10-1)  & 218.9 & 531 \\
QTRL-5 & QNN: ($U_3$ block)*5 + Mapping Model: (11-10, 10-10, 10-1)  & 493.8 & 651 \\
\hline
\end{tabular}
\caption{
Comparative performance of Classical Policy and QT models with varying numbers of QNN blocks for \textsf{\small CartPole-v1} environment.}
\label{table:cartpole}
\end{table*}

\begin{table*}[!htp]
\centering
\begin{tabular}{|l|l|c|c|}
\hline
\textbf{Model} & \textbf{Topology} & \makecell{\textbf{Last 10 episode} \\ \textbf{average reward}} & \textbf{Para. size} \\
\hline
Classical Policy Model & (147-32, 32-3) & 0.916 & 4835 \\
\hline
QTRL-3 & QNN: ($U_3$ block)*3 + Mapping Model: (11-10, 10-10, 10-1) & 0.694 & 1749 \\
QTRL-7 & QNN: ($U_3$ block)*7 + Mapping Model: (11-10, 10-10, 10-1)  & 0.827 & 2061 \\
QTRL-13 & QNN: ($U_3$ block)*13 + Mapping Model: (11-10, 10-10, 10-1)  & 0.900 & 2529 \\
\hline
\end{tabular}
\caption{
Comparative performance of Classical Policy and QT models with varying numbers of QNN blocks for \textsf{\small MiniGrid-Empty-5x5-v0} environment.}
\label{table:minigrid}
\end{table*}

Beginning with the \textsf{\small CartPole-v1} environment, as shown in Table~\ref{table:cartpole}, the classical policy model used consists of 898 parameters. This implies that the corresponding required qubits for QTRL is $\lceil \log_2 898 \rceil = 10$. We trained the model for 2000 episodes, and as shown in Fig.~\ref{fig:scheme}(b), the QTRL with varying depths $(L = 1, 3, 5)$ are depicted in different shades of blue, while the classical method is shown in red. The deeper circuits, having better expressibility, result in the total reward curve being closer to the classical case. As emphasized in Table~\ref{table:cartpole}, in the $L = 1$ case, the 898 classical policy model parameters are generated by only 361 QNN and mapping model parameters, while in the $L = 3$ case, 531 parameters are used in the QNN and mapping model. In the $L = 5$ case, we found that the last 10 episode average reward, 493.8, is better than the classical case of 436.3, while using only 651 parameters during training.

In the \textsf{\small MiniGrid-Empty-5x5-v0} environment, we observed similar behavior to the results in \textsf{\small CartPole-v1}, although \textsf{\small MiniGrid-Empty-5x5-v0} is a slightly more complicated environment.  We trained the model for 4000 episodes. As shown in Table~\ref{table:minigrid}, the classical policy model utilizes 4835 parameters, mainly due to the larger size of the observation space in this environment, and implying the qubit counts for QTRL is $\lceil \log_2 4835 \rceil = 13$. In Fig.~\ref{fig:scheme}(c), the classical method is shown in red, while QTRL with varying depths $(L = 3, 7, 13)$ are shown in different shades of blue. As noted in Table~\ref{table:minigrid}, while shallower circuits achieve lower average rewards, it is worth noting that in the $L=13$ case, we achieved a last 10 episode average reward of 0.900 with only 2529 QNN and mapping model parameters, whereas the classical approach with an average reward of 0.916 requires 4835 parameters.

The results from our experiments on the \textsf{\small CartPole-v1} and \textsf{\small MiniGrid-Empty-5x5-v0} environments demonstrate the potential of the QTRL approach to effectively train reinforcement learning policies using fewer parameters than classical methods. This is particularly evident in the deeper quantum circuits, which show performance close to or even surpassing classical methods while utilizing significantly fewer parameters. 

One may observe in Fig.~\ref{fig:scheme}(a) that the QNN part only consists of the $U_3$ ansatz without data encoding layers, unlike conventional QML and QRL approaches \cite{dataru1, qtl1, qrl_ec1, qrl1, qrl2}. This is because our QNN only serves to generate parameters for the classical policy model, meaning that both the observation state input and the action output are handled in the classical part. Without the data encoding issue, the circuit width and depth are independent of the input data size, making our approach extremely practical for dealing with larger observation state inputs. In conventional approaches, some data might be too large to construct a reasonable circuit either in depth or width (qubit counts).

Continuing with the classical data input and output features of QTRL, this implies that once we use the training flow in Fig.~\ref{fig:scheme}(a), the training result is the generated classical policy model $\pi_\theta$, which is identical to any other classical model trained in a conventional manner. This means that:
\begin{enumerate}
    \item This approach is compatible with any potential transfer learning or fine-tuning on the classical side. We can use the quantum computer to initially train the model with fewer parameters and then use a classical computer for full parameter fine-tuning.
    \item \label{p2} The model is now decoupled from quantum computing hardware. Thus, in the inference stage, this ``quantum trained'' model can be run entirely on classical computing hardware.
\end{enumerate}

Point (\ref{p2}) is particularly important in RL tasks because these tasks usually demand low-latency feedback from the policy model. For example, in Autonomous Driving tasks \cite{autocar1}, using conventional QRL or hybrid quantum-classical RL (QCRL) approaches, where quantum jobs might need to be submitted to a cloud quantum computing platform and queued, is unlikely to meet the real-time response requirements of the fast-changing road environment. In contrast, our QTRL approach produces a classical model that can be executed on the vehicle's edge device. Similar situations may arise in tasks interacting with real-time data in the real world. This demonstrates the widespread potential of our QTRL approach in terms of practicality. 

Furthermore, considering the current and likely future stages of quantum computing development, the financial cost of using quantum computing hardware is much higher than using classical CPUs and GPUs. The classical inference feature of QTRL is much more cost-efficient compared to QRL and QCRL approaches, which require quantum computer access every time the model is used. This cost efficiency further enhances the practicality and appeal of the QTRL approach.

\section{Conclusion}
\label{sec:conclusion}
In this work, we introduced the QTRL approach, which leverages QNN to generate parameters for classical policy models. This method addresses significant challenges in QRL and QML, such as data encoding and dependency on quantum hardware during inference. Our experiments on the \textsf{\small CartPole-v1} and \textsf{\small MiniGrid-Empty-5x5-v0} environments demonstrate that QTRL can achieve comparable or superior performance to classical methods while using significantly fewer parameters. This efficiency makes QTRL particularly suitable for RL tasks requiring low-latency feedback and practical for real-world applications like autonomous driving. Moreover, QTRL's independence from quantum hardware during inference and its compatibility with classical transfer learning and fine-tuning techniques enhance its practicality and cost-efficiency. This makes QTRL a promising approach for various applications, offering both practical benefits and theoretical advancements in reinforcement learning. 

QTRL not only bridges the gap between quantum and classical RL but also provides a practical solution that can be readily applied with existing classical infrastructure. This approach paves the way for more accessible and scalable RL models, harnessing the strengths of both quantum and classical computing paradigms.



\bibliographystyle{IEEEtran}
\bibliography{references}

\end{document}